\documentclass[floatfix,twocolumn]{aastex63}
\def\aap{{A\&A}}                % Astronomy and Astrophysics
\usepackage{graphicx}
\usepackage{amsmath}
\usepackage{amssymb}
\usepackage{color}
\usepackage{diagbox}

\def\apjl{ApJ Letters}
\def\apjs{ApJ Supplements}

\begin{document}
\title{Supermassive Binary Black Hole Evolution\\
can be traced by a small SKA Pulsar Timing Array
}
\date{\today}

\correspondingauthor{Yi Feng}
\email{yifeng@nao.cas.cn}
\author{Yi Feng}
\affil{CAS Key Laboratory of FAST, National Astronomical Observatories, Chinese Academy of  Sciences, Beijing 100101, People's Republic of China}
\affil{University of Chinese Academy of Sciences, Beijing 100049, People's Republic of China}
\affil{CSIRO Astronomy and Space Science, PO Box 76, Epping, NSW 1710, Australia}

\author{Di Li}
\affil{CAS Key Laboratory of FAST, National Astronomical Observatories, Chinese Academy of  Sciences, Beijing 100101, People's Republic of China}
\affil{University of Chinese Academy of Sciences, Beijing 100049, People's Republic of China}

\author{Zheng Zheng}
\affil{CAS Key Laboratory of FAST, National Astronomical Observatories, Chinese Academy of  Sciences, Beijing 100101, People's Republic of China}
\author{Chao-Wei Tsai}
\affil{CAS Key Laboratory of FAST, National Astronomical Observatories, Chinese Academy of  Sciences, Beijing 100101, People's Republic of China}

% Mark off the abstract in the ``abstract'' environment. 
\begin{abstract}
Supermassive  black holes are commonly found in the center of galaxies and evolve with their hosts. The supermassive binary black holes (SMBBH) are thus expected to exist in close galaxy pairs, however, none has been unequivocally detected. The square kilometre array (SKA) is a multi-purpose radio telescope with a collecting area approaching 1 million square metres, with great potential for detecting nanohertz gravitational waves (GWs). In this paper, we quantify the GW detectability by SKA for a realistic SMBBH population using pulsar timing array (PTA) technique and quantify its impact on revealing SMBBH evolution with redshift for the first time. With only $\sim20$ pulsars, much smaller a requirement than in previous work, the SKA PTA is expected to obtain detection within about 5 years of operation and to achieve a detection rate of more than 100 SMBBHs/yr after about 10 years. Although beyond the scope of this paper, we must acknowledge that the presence of persistent red noise will reduce the number of expected detections here. It is thus imperative to understand and mitigate red noise in the PTA data.
The GW signatures from a few well-known SMBBH candidates, such as OJ 287, 3C 66B, NGC 5548 and Ark 120, will be detected given the currently best-known parameters of each system. 
Within 30 years of operation, about 60 individual SMBBHs detection with $z<0.05$ and more than $10^4$ with $z<1$ are expected. The detection rate drops precipitately beyond $z=1$. The substantial number of expected detections and their discernible evolution with redshift by SKA PTA will make SKA a significant tool for studying SMBBHs.

\end{abstract}

\pacs{abcd}
\keywords{pulsar timing array: general --- continuous gravitational waves: detection algorithm}

\section{Introduction}
GW observatories such as advanced LIGO (aLIGO) \citep{2015CQGra..32g4001L} and Virgo \citep{2015CQGra..32b4001A} have reached remarkable sensitivities in the high frequency band ($\sim 10-1000\,\rm Hz$). 
The detection of GWs from compact binary mergers has become a regular occurrence.
In the nanohertz frequency band, pulsar timing arrays (PTAs), in which a collection of millisecond pulsars is monitored, can be used to detect and study GWs \cite{Detweiler1979,Foster_Backer90}.  
The primary single source of GWs in nanohertz band are believed to be inspiralling SMBBHs, formed in the aftermath of galaxy mergers \citep{1980Natur.287..307B}. 
The detection of SMBBH systems can yield direct information about the masses and spins of the black holes \citep{2012PhRvL.109h1104M}. These single GW sources can also be studied by coordinated electromagnetic observations, thus enabling a multi-messenger view of the black hole systems \cite{2013CQGra..30v4013B,2019BAAS...51c.490K}. 

PTA based GW astronomy is expected to progress significantly with the new and high sensitivity radio telescopes such as FAST \citep{2011IJMPD..20..989N} and SKA \citep{2009A&A...493.1161S}. 
The pulsar timing effort by SKA will significantly enhance the sensitivity of the current PTA networks by providing a larger number of newly discovered millisecond pulsars (MSPs), and better timing precision on the existing and new MSPs.
Upon its completion, SKA will be the most sensitive telescope for detecting nanohertz GWs in the next generation. Thus, it is important to estimate the GW detection abilities of SKA. 
Wang \& Mohanty \citep{2017PhRvL.118o1104W} carried out a pioneering quantitative assessment of the GW detectability for individual SMBBH searches with a simulated SKAPTA containing $10^3$ pulsars and found the SKAPTA to significantly increase the maximum distance of detectable GWs emitted by SMBBHs. 
They also considered two realistic candidates, namely, PG 1302-102 and PSO J334+01. However, no previous work has provide actual predictions of the number of SMBBHs to be detected by SKAPTA and their evolution through redshifts.

Here we provide one of the first quantitative estimates on the GW detectability for a realistic individual SMBBH population. We tackle the problem in three steps. First, a SKAPTA detection curve, namely the minimum detectable GW strain amplitude as a function of GW frequency, is
calculated. A SKAPTA containing 20 randomly placed pulsars with timing root mean square (RMS) of 20\,ns is used to compute the detection curves, using the $\mathcal{F}_{e}$ statistic, the logarithm of the likelihood ratio maximized over the signal parameters developed by \cite{1998PhRvD..58f3001J, 2012ApJ...756..175E}. Here we assume circular binary orbits for the SMBBHs.
Second, an expected SMBBH population in the PTA frequency band is constructed based on the probability of a galaxy hosting a SMBBH in the PTA band and the population of host galaxies. 
The SMBBH population is estimated following the descriptions of \cite{2017NatAs...1..886M,chiara_mingarelli_2017_838712} (hereafter M17). 
M17 used data from local galaxies in the 2 Micron All Sky Survey (2MASS) \citep{2006AJ....131.1163S} Extended Source Catalog \citep{2000AJ....119.2498J} and galaxy merger rates from Illustris \cite{2015MNRAS.449...49R,2014MNRAS.445..175G}.
Due to the expected great improvement brought about by SKAPTA, we much extend the redshift range in M17 using galaxy stellar mass function from \cite{2003ApJS..149..289B,2013ApJ...777...18M} (GSMF). 
Third, we extract detectable GW sources within each redshift bin according to the SKAPTA detection curve calculated above and the expected SMBBH population estimated in step two. Our recipe allows for a quantitative prediction of the properties of the SMBBH population to be detected by SKAPTA, such as the number of detections per redshift bin and detection rates per year, for the first time. 
%%%%%%%%%%%%%%%%%%%%%%%%%%%%%%%%%%%%%%%%%%%%%%%%%%%%%%%%%%%%%%%%%%%%%%%%

\section{Simulated SKAPTA}
With broad frequency bands and massive collecting areas, the radiometer noise for some of the brightest pulsars can be reduced from current 100\,ns level down to below 10\,ns by the large radio telescopes like FAST and SKA. Jitter noise, which is assumed to be caused by the fluctuation in the shape and arrival time of individual pulses, will limit the timing precision achievable over data spans of a few years for these large facilities \citep{2019RAA....19...20H}. 
To estimate the timing RMS, we assume that the timing RMS is white and consists solely of radiometer noise and jitter noise. We defer the influence of red noise (e.g. GWB, dispersion measure variation noise, intrinsic timing noise \citep{2019RAA....19...20H}) to later discussions. Table 4 in \citep{2018PhRvD..98j2002P} listed white noise for 10 Parkes Pulsar Timing Array (PPTA) \citep{2013PASA...30...17M} pulsars with a harmonic mean of 20\,ns for integration time of 30 minutes with the SKA Phase 1\citep{2015aska.confE..37J}. Within the SKA sky, International Pulsar Timing Array (IPTA)\citep{2013CQGra..30v4010M} source list contains more millisecond pulsars than PPTA does. For example, IPTA PSR~J0023+0923, PSR~J0030+0451, PSR~J0931$-$1902 are not in the PPTA line-up. Considering the better sensitivity of the full SKA  than that of SKA Phase 1, which will further improve timing RMS, we assume a conservative 20 pulsars with a harmonic mean of 20\,ns. We thus construct a mock SKAPTA data set containing 20 millisecond pulsars randomly distributed in the sky. Noise realizations are drawn from an independent and identically distributed (i.i.d.) $\mathcal{N}(0,\sigma^2)$ (zero mean white Gaussian noise) process, 
with $\sigma = 20$~ns for all pulsars. 
We choose the cadence to be 20 $\rm{yr}^{-1}$ in order to match the typical cadence used in current PTAs. 
%
%%%%%%%%%%%%%%%%%%%%%%%%%%%%%%%%%%%%%%%%%%%%%%%%%%%%%%%%%%%%%%%%%

\section{SMBBH population in the PTA band}
The SMBBH population emitting in the PTA frequency band depends on two quantities:
\begin{enumerate}
\item {\it The probability of a galaxy hosting a SMBBH in the PTA band}. We exploit the approach put forward by M17 (See M17 for details of the approach).
The probability is the multiplication of the probability that a SMBBH is in the PTA band and the probability that a galaxy hosts a SMBBH. The probability that a SMBBH is in the PTA band is 
$t_\mathrm{obs}/T_\mathrm{life}$, where $t_\mathrm{obs} = 5/256 c^5 (\pi f)^{-8/3} {[G M_c(1+z)]}^{-5/3}$ is the time to coalescence of the binary in the observed frame \citep{1964PhRv..136.1224P}.
Here $f=1\,\mathrm{nHz}$, chirp mass $M_c = \left[ q/(1+q)^2 \right]^{3/5} M_\bullet$ with black hole mass ratio $q$ drawn from a log-uniform distribution in $[0.25,1]$. 
The SMBBH total mass $M_\bullet$ is estimated using the  $M_\bullet-M_{\mathrm{bulge}}$ empirical scaling relation from \citep{2013ApJ...764..184M}. 
As discussed in M17, only massive early-type galaxies are considered in this simulation, therefore we take the galaxy stellar mass $M_*$ as $M_{\mathrm{bulge}}$ for $M_\bullet$ estimates.
$T_\mathrm{life}$ is the effective lifetime of the binary, which is the sum of the dynamical friction ($t_\mathrm{df}$) \citep{BinneyTremaine} and stellar hardening ($t_\mathrm{sh}$) \citep{2015MNRAS.454L..66S} timescales.
The probability that a galaxy hosts SMBBH is computed using the Illustris \cite{2015MNRAS.449...49R,2014MNRAS.445..175G} cumulative galaxy-galaxy merger rate, $\mathrm{d}N/\mathrm{d}t(M_*, z', \mu_*)$ where $\mu_*$ is the stellar mass ratio of the galaxies, taken at the beginning of the binary evolution at redshift $z'$, which is calculated at lookback time of $T_\mathrm{life}+T_\mathrm{lookback}$ with Planck cosmological parameters \citep{2016A&A...594A..13P}, here $T_\mathrm{lookback}$ is the lookback time at $z$.
To summarize, probability of a galaxy hosting a SMBBH in the PTA band is:
\begin{equation}
\label{eq:prob}
p= \frac{t_\mathrm{obs}}{T_\mathrm{life}}\int_{0.25}^{1} \mathrm{d}\mu_* \frac{\mathrm{d}N}{\mathrm{d}t}(M_*, z', \mu_*) T_\mathrm{life}\, ,    
\end{equation}
\item {\it The population of host galaxies}. As discussed in M17, we only consider massive early-type galaxies with galaxy stellar mass greater than $10^{11}\,\rm M_{\odot}$. In addition, we impose a cut on the galaxy population at galaxy stellar mass $M_*<10^{12}\,\rm M_{\odot}$ because such massive galaxies are rare. 
For a population of galaxy with known redshift $z$ and $M_*$, we can calculate the probability of a selected galaxy hosting a SMBBH in the PTA band using Eq. \ref{eq:prob}, and thus determining the population of SMBBH emitting in the PTA frequency band. For galaxies at $z<0.05$, we use the galaxy catalog in M17, which selected galaxies at $z<0.05$ from the 2 Micron All Sky Survey (2MASS) \citep{2006AJ....131.1163S} Extended Source Catalog \citep{2000AJ....119.2498J}. 
To approximate a mass selection for more distant galaxies, we use GSMF given in Table 4 of \citep{2003ApJS..149..289B} and Table 1 of  \citep{2013ApJ...777...18M} for redshift interval between $z$ = 0.05, 0.2, 0.5, 1.0, 1.5, 2.0, 2.5, 3.0, and 4.0. For each $z$ interval at $z>0.05$, we randomly choose $10^6$ host galaxies with $z$ drawn from a uniform distribution and stellar mass drawn from the corresponding GSMF. 
The number of $10^6$ galaxies is used to ensure stable simulation results in the Monte Carlo process. The estimated numbers of galaxies in each redshift bin is then scaled to the expected galaxy population from the sample of $10^6$ galaxies in the simulation. 
Using Eq. \ref{eq:prob}, we calculate the probability of a selected galaxy hosting a SMBBH in the PTA band. We then generate a random number from $U$[0, 1], if the random number is smaller than the probability, then the galaxy is considered to host a true SMBBH. Finally, the inclination and polarization-averaged strain and GW frequency of the SMBBH are calculated as described in M17 using 
\begin{equation}
\label{eq:h}
h=\sqrt{\frac{32}{5}}\frac{{M}_c^{5/3}}{D_c} \left[\pi f(1+z)\right]^{2/3}\, , 
\end{equation}
\begin{equation}
\label{eq:freq}
f=\pi^{-1}\left[\frac{G{M}_c(1+z)}{c^3}\right]^{-5/8}\left[ \frac{256}{5}(t_\mathrm{obs}-t)\right]^{-3/8} \, ,
\end{equation}
where $D_c$ is the comoving distance of the binary, $(t_\mathrm{obs}-t)$ is drawn from a uniform distribution in $[26~\rm{Myr}, 100~\rm{yr}]$.
\end{enumerate}
%%%%%%%%%%%%%%%%%%%%%%%%%%%%%%%%%%%%%%%%%%%%%%%%%%%%%%%%%%%%%%%%%%%
\begin{figure}
%\centering
\begin{center}
\includegraphics[width=0.5\textwidth]{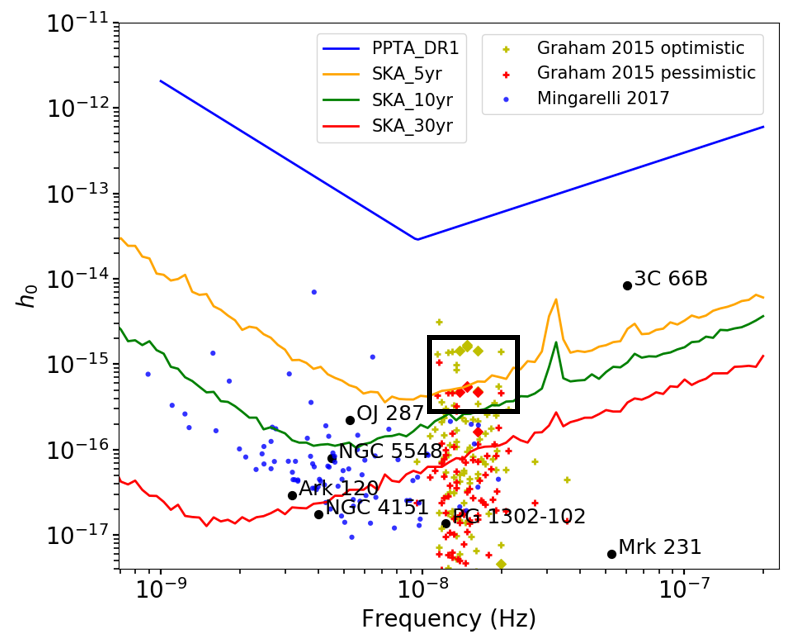}
\caption{Detection curves for PPTA\_DR1 (blue line), SKA\_5yr (orange line), SKA\_10yr (green line), SKA\_30yr (red line) respectively. The seven black dots represent the SMBBH candidates discussed in \citep{2019arXiv190703460F}. The yellow (best case) and red (pessimistic case) dots ('+' for $z<2$ samples and diamond symbol for $z>2$) represent CRTS samples. The diamond symbols in the black box represent 4 unreliable candidates (two overlapped, see the text for details). The blue dots represent 87 single GW sources (one realization of sky) from M17.}
\label{pop1}
%'D:\python_code\ska/sensitivity_curve_dili_20190527.py', where get pta gain
\end{center}
\end{figure}
%%%%%%%%%%%%%%%%%%%%%%%%%%%%%%%%%%%%%%%%%%%%%%%%%%%%%%%%%%%%%%%%%%%

\section{Results}
We use the $\mathcal{F}_{e}$ statistic with False Alarm Probability of $10^{-3}$ to calculate detection curves for SKAPTA with total time span of 5, 10, 30\,yrs respectively. The results are shown in Figure \ref{pop1}.
The SMBBH candidates such as 3C 66B \citep{Iguchi2010}, OJ 287 \citep{Valtonen2016}, NGC 5548 \citep{2016ApJ...822....4L} and Ark 120 \citep{2019ApJS..241...33L} (black dots in Figure 1) discussed in \citep{2019arXiv190703460F} can be detected if they are true SMBBHs.
For the other SMBBH candidates Mrk 231 \citep{2015ApJ...809..117Y}, PG 1302-102 \citep{2015Natur.518...74G}, NGC 4151 \citep{2012ApJ...759..118B}, they are hard to detect even in SKA era for their weaker estimated GW signature caused by their small chirp masses.
\citep{2015MNRAS.453.1562G} proposed 111 SMBBH candidates by inspecting the light curves of $\sim$250\,k quasars identified in the Catalina Real-time Transient Survey (CRTS, \cite{2009ApJ...696..870D}). 
We plot the 98 candidates (hereafter CRTS samples) with reported black hole mass estimates for optimistic case assuming mass ratio $q = 1$ (yellow dots in Figure 1) and pessimistic case assuming $q = 0.1$ (red dots in Figure 1). At least 10 single sources in CRTS samples can be detected, assuming these are all true sources, even for the pessimistic case. Unfortunately, CRTS samples are likely contaminated by several false positives \cite{2018ApJ...856...42S} and we will discuss this later. 
The blue dots represent 87 single GW sources (one realization of sky) from M17. 0, 2, 12, 64 sources of M17 can be detected for PPTA\_DR1, SKA\_5yr, SKA\_10yr, SKA\_30yr. \\
\begin{figure}
%\centering
\begin{center}
\includegraphics[width=0.5\textwidth]{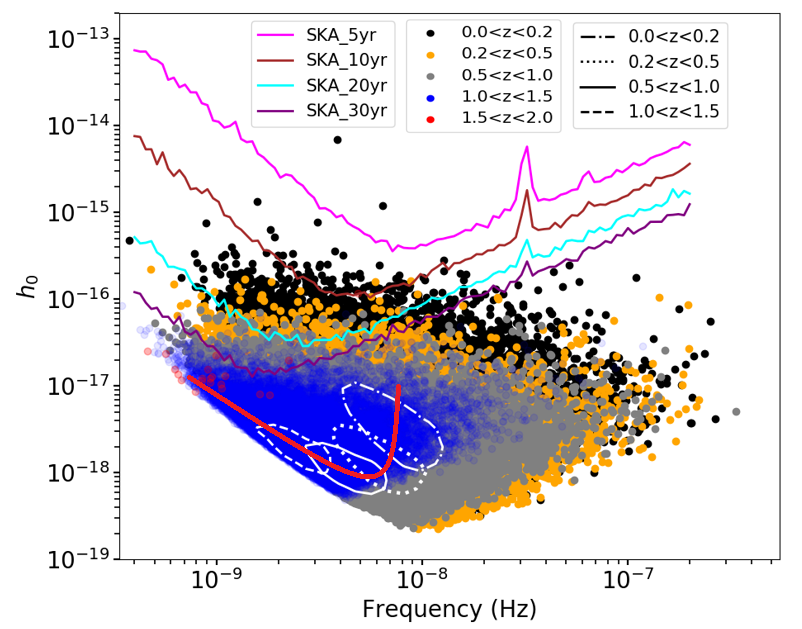}
\caption{Detection curves for SKA\_5yr (magenta line), SKA\_10yr (brown line), SKA\_20yr (cyan line), SKA\_30yr (purple line) respectively. Black, orange, gray, blue, red dots represent SMBBH population hosted by $10^6$ galaxies from $0.0<z<0.2$, $0.2<z<0.5$, $0.5<z<1.0$, $1.0<z<1.5$, $1.5<z<2.0$ respectively (for $0.0<z<0.2$, it is 87 SMBBHs from M17 combined with SMBBH population hosted by $10^6$ galaxies from $0.05<z<0.2$). Dashdot, dotted, solid, dashed number density contours represent 50\% of the peak value for $0.0<z<0.2$, $0.2<z<0.5$, $0.5<z<1.0$, $1.0<z<1.5$ respectively. For $z>2.0$, there are no detectable SMBBHs in the PTA band, so SMBBH population at $z>2.0$ is not shown in the figure. The red curve cross the contour centers shows the evolution trend of SMBBH population from low redshift to high redshift.}
\label{pop2}
%D:D:\python_code\ska/visual.py
\end{center}
\end{figure}
Figure \ref{pop2} shows the SMBBH population in the PTA band. 
In Figure \ref{pop2}, we also plot the 5, 10, 20, 30\,yr detection curves to show the potential detectability of these single GW sources. 
The strain of the population shifts to a low value at $z\lesssim0.5$ for higher redshift due to their further distance.  
The frequency of the population shifts to lower frequency at $z\gtrsim0.5$ because these SMBBHs do not have enough time to evolve to be closer, with higher redshift SMBBHs have less time to evolve. 
Given the detection curves and SMBBH population in the PTA band, we determine the detection number of single GW sources as a function of SKAPTA time span. 
The source is considered detected if it lies above the detection curve.
The total number of host galaxies is the multiplication of comoving volume and integral of the corresponding GSMF from $10^{11}\,\rm M_{\odot}$ to $10^{12}\,\rm M_{\odot}$, and is listed in the last row of Table \ref{tab:number}. The number of SMBBHs in the PTA band is listed in the second last row. 

\begin{figure}
%\centering
\begin{center}
\includegraphics[width=0.5\textwidth]{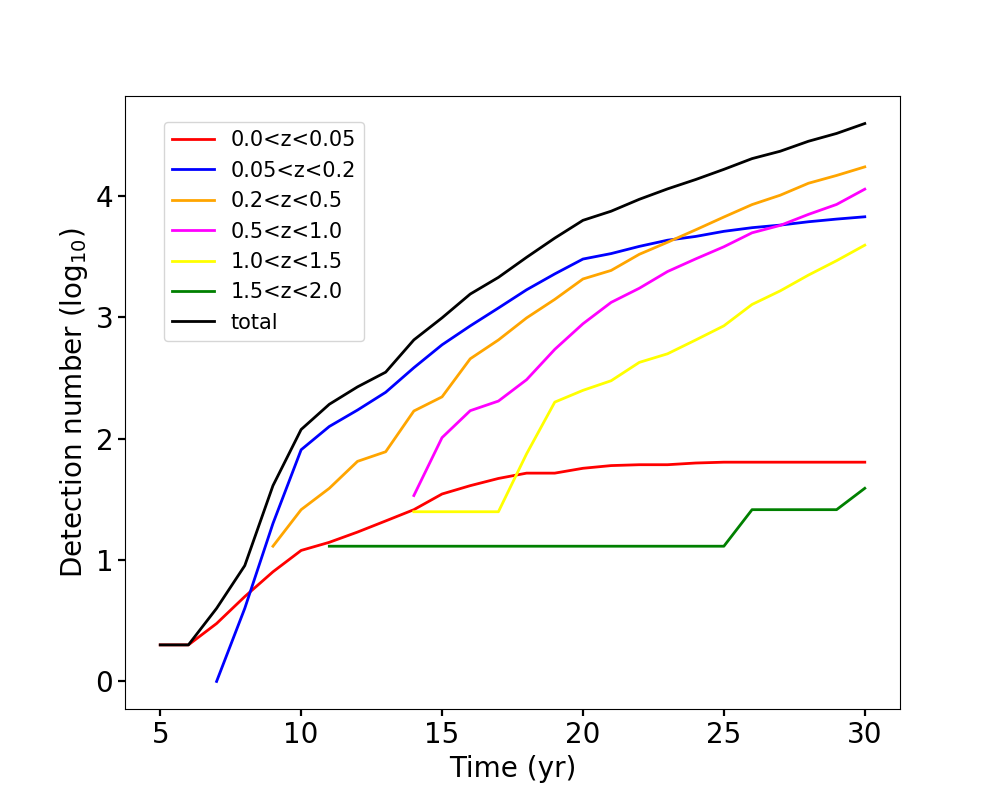}
\caption{Detection number of single GW sources versus time span of SKAPTA. The red, blue, orange, magenta, yellow, green, black colors represent number of sources at $0.0<z<0.05$, $0.05<z<0.2$, $0.2<z<0.5$, $0.5<z<1.0$, $1.0<z<1.5$, $1.5<z<2.0$ and total number respectively.}
\label{fig:number}
%D:\python_code\ska\detection_number.py
\end{center}
\end{figure}

\begin{figure}
%\centering
\begin{center}
\includegraphics[width=0.5\textwidth]{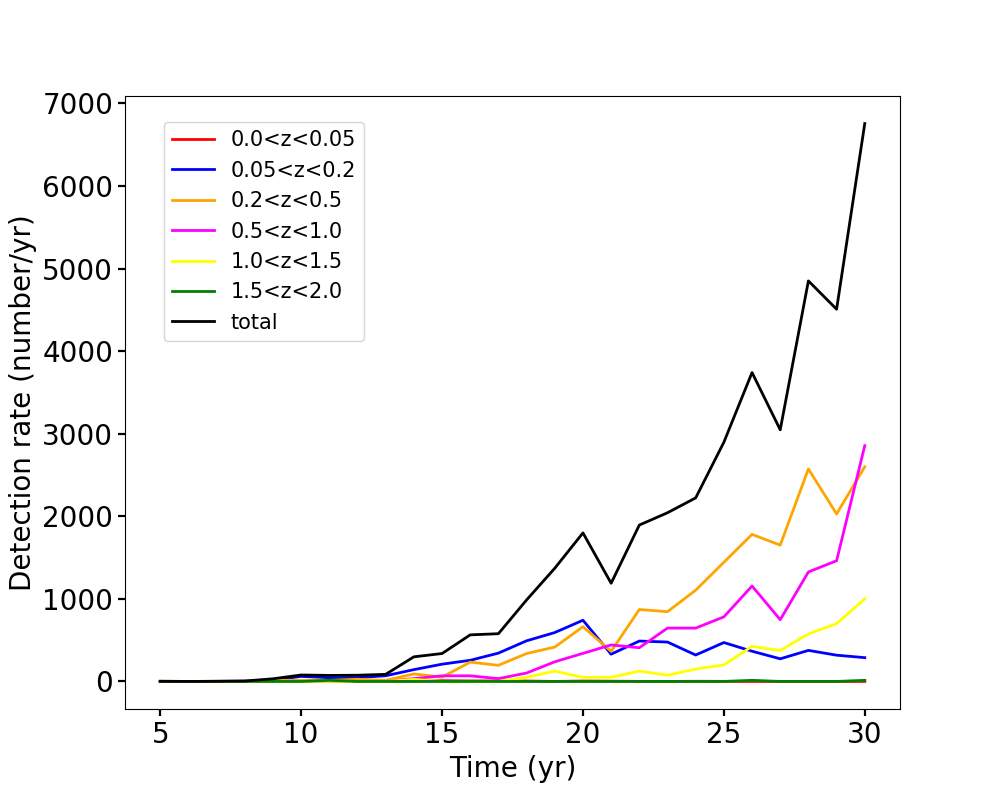}
\caption{Detection rate of single GW sources versus time span of SKAPTA. The red, blue, orange, magenta, yellow, green, black colors represent number of sources at $0.0<z<0.05$, $0.05<z<0.2$, $0.2<z<0.5$, $0.5<z<1.0$, $1.0<z<1.5$, $1.5<z<2.0$ and total rate respectively.}
\label{fig:rate}
%D:\python_code\ska\detection_rate.py
\end{center}
\end{figure}

Figure \ref{fig:number} shows detection number for different redshift ranges as a function of SKAPTA time span. We list the detection number for different redshift ranges of 5, 10, 15, 20, 30 time span SKAPTA in Table \ref{tab:number}. 
More than $10^4$ sources can be detected by SKAPTA after 30 years of operation. The primary detectable sources come from galaxies at $z<1$.
Detectable sources at $z>1.5$ are rare as their larger distances dampen the apparent GW signal. Some of them also did not have enough time to evolve to be close enough, thus have a longer observed orbital period outside of the PTA frequency band. The dramatic drops of GW detection at $z>1.5$ is consistent with \cite{2013MNRAS.433L...1S, 2015MNRAS.447.2772R}. Moreover, no hosts at $z>2.0$ are expected to have detectable SMBBHs in our simulation. If this result is true, it implies that CRTS samples at $z>2.0$ may not be real SMBBHs.
For example, Table 1 in \cite{2018ApJ...856...42S} listed top 10 candidates in CRTS samples providing the largest contribution to the expected GWB, which are likely to be false positives. 
In this sample, four of them (i.e., HS~0926+3608, SDSS~J140704.43+273556.6, SDSS~J131706.19+271416.7, SDSS~J134855.27-032141.4) have redshifts $z>2.0$ and are shown using diamond symbols inside the black box in Figure \ref{pop1}. Combining our results and \cite{2018ApJ...856...42S}, these 4 candidates can be unreliable.\\
\begin{table*}
\begin{center}
\caption{detection number}
\begin{tabular}{|c|c|c|c|c|c|c|}
\hline %\hline
\diagbox{time(yr)}{redshift} & $0.0<z<0.05$ & $0.05<z<0.2$ & $0.2<z<0.5$& $0.5<z<1.0$& $1.0<z<1.5$ & $1.5<z<2.0$\\
\hline
5 &  2 & 0   &   0  &0&0 &0\\
10 &  12 & 81   &   26  &0&0 &0\\
15 &  35 & 593   &   221  &102&25 &13\\
20 &  57 & 3017   &   2067  &884&250 &13\\
30 &  64 & 6726   &   17329  &11356&3925 &39\\
\hline
total SMBBHs &  87 & $1.0\times10^5$   & $1.0\times10^6$  &$3.4\times10^6$ &$8.4\times10^5$ &$2.6\times10^2$\\
\hline
total hosts &  5119 & $1.6\times10^6$   &   $1.3\times10^7$  &$3.4\times10^7$&$2.5\times10^7$&$1.3\times10^7$\\
\hline %\hline
\end{tabular}
\label{tab:number}
\end{center}
%\tablecomments{haha}
\end{table*}
%------------------------------------------------------------------
We calculate the detection rate as a function of SKAPTA time span by comparing the detection number of two consecutive years. 
The results are shown in Figure \ref{fig:rate}. Unlike GW sources in LIGO frequency band, the detection rate of single GW sources of SKAPTA is not uniform in time. The detectability increases slowly in the early times, but increases faster to be more than 100 detections/yr after about 10 yrs. 
This is a stipend from the unique PTA based search, which accumulates SNR with time, highlighting the importance of long time span of a PTA campaign.  
%%%%%%%%%%%%%%%%%%%%%%%%%%%%%%%%%%%%%%%%%%%%%%

\section{Discussion}
Red noise was ignored under the assumption that it can be mitigated to a low noise level or it does not influence single GW detection using special techniques. If the timing residuals have a strong red noise component emulating an unresolved GWB with amplitude of $4\times 10^{-16}$, the detection number decreases drastically as shown in Figure 5. 
This is because PTA is less sensitive to high frequency ($> 1\,\rm{yr}^{-1}$) SMBBHs, and strong red noise in timing residuals greatly diminishes the chance of detecting lower frequency SMBBHs. The SKAPTA GW detections depend on how well the GWB can be subtracted, which should be further studied. 
The methodology of GWB subtraction is also important to mitigate other various types of red noises which could limit the detectability of SMBBHs with SKAPTA.  
\begin{figure}
%\centering
\begin{center}
\includegraphics[width=0.5\textwidth]{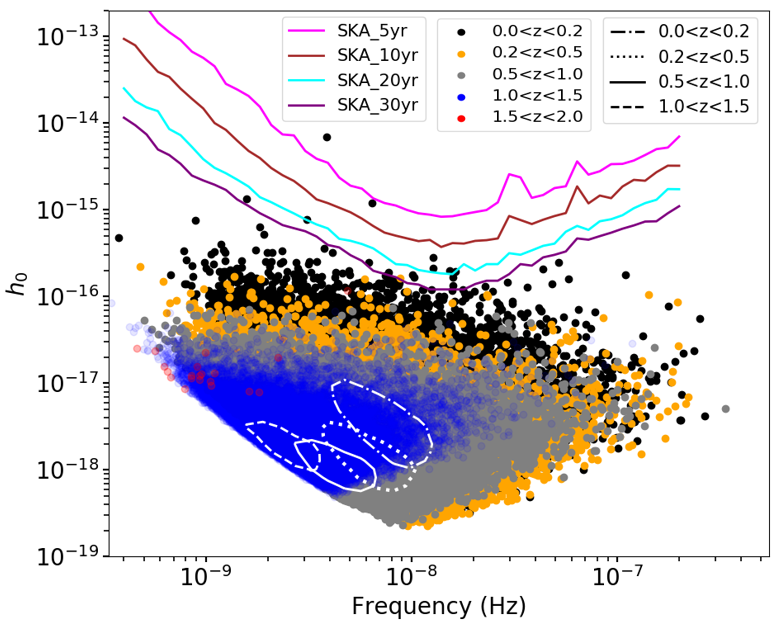}
\caption{Same as Fig. \ref{pop2}, but timing residuals have a strong red noise component of $4\times 10^{-16}$, emulating the amplitude of an unresolved GWB.}
\label{fig:red}
%D:\python_code\ska\detection_rate.py
\end{center}
\end{figure}
%%%%%%%%%%%%%%%%%%%%%%%%%%%%%%%%%%%%%%%%%%%%%%

\section{Conclusion}
The unprecedented sensitivity of SKA facilitates realization of a significant PTA with a small number of pulsars. The SKAPTA in this work consists of only 20 pulsars versus $\sim1000$ in previous works. Such a simple SKAPTA can still detect a large number of SMBBHs and enable studies of their evolution through redshift up to $z = 1-2$ assuming red noise in the PTA data can be mitigated. The presence of red noise will reduce the number of detectable individual sources. Nevertheless, SKAPTA will be a revolutionary instrument for studying SMBBH evolution.

\begin{acknowledgments}
We thank the anonymous referee for very useful comments on the manuscript. This work is supported by National Natural Science Foundation of China (NSFC) programs, No. 11988101, 11725313, 11690024, 11703036, by the CAS International Partnership Program No.114-A11KYSB20160008, by the CAS Strategic Priority Research Program No. XDB23000000 and the National Key R\&D Program of China (No. 2017YFA0402600).

\end{acknowledgments}

\bibliographystyle{aasjournal}
\bibliography{haha}

\begin{thebibliography}{}
\expandafter\ifx\csname natexlab\endcsname\relax\def\natexlab#1{#1}\fi
\providecommand{\url}[1]{\href{#1}{#1}}
\providecommand{\dodoi}[1]{doi:~\href{http://doi.org/#1}{\nolinkurl{#1}}}
\providecommand{\doeprint}[1]{\href{http://ascl.net/#1}{\nolinkurl{http://ascl.net/#1}}}
\providecommand{\doarXiv}[1]{\href{https://arxiv.org/abs/#1}{\nolinkurl{https://arxiv.org/abs/#1}}}

\bibitem[{{Acernese} {et~al.}(2015){Acernese}, {Agathos}, {Agatsuma}, {Aisa},
  {Allemandou}, {Allocca}, {Amarni}, {Astone}, {Balestri}, \&
  {Ballardin}}]{2015CQGra..32b4001A}
{Acernese}, F., {Agathos}, M., {Agatsuma}, K., {et~al.} 2015, Classical and
  Quantum Gravity, 32, 024001, \dodoi{10.1088/0264-9381/32/2/024001}

\bibitem[{{Begelman} {et~al.}(1980){Begelman}, {Blandford}, \&
  {Rees}}]{1980Natur.287..307B}
{Begelman}, M.~C., {Blandford}, R.~D., \& {Rees}, M.~J. 1980, \nat, 287, 307,
  \dodoi{10.1038/287307a0}

\bibitem[{{Bell} {et~al.}(2003){Bell}, {McIntosh}, {Katz}, \&
  {Weinberg}}]{2003ApJS..149..289B}
{Bell}, E.~F., {McIntosh}, D.~H., {Katz}, N., \& {Weinberg}, M.~D. 2003, \apjs,
  149, 289, \dodoi{10.1086/378847}

\bibitem[{{Binney} \& {Tremaine}(2008)}]{BinneyTremaine}
{Binney}, J., \& {Tremaine}, S. 2008, Galactic Dynamics: Second Edition
  (Princeton University Press)

\bibitem[{{Bon} {et~al.}(2012){Bon}, {Jovanovi{\'c}}, {Marziani},
  {Shapovalova}, {Bon}, {Borka Jovanovi{\'c}}, {Borka}, {Sulentic}, \&
  {Popovi{\'c}}}]{2012ApJ...759..118B}
{Bon}, E., {Jovanovi{\'c}}, P., {Marziani}, P., {et~al.} 2012, \apj, 759, 118,
  \dodoi{10.1088/0004-637X/759/2/118}

\bibitem[{{Burke-Spolaor}(2013)}]{2013CQGra..30v4013B}
{Burke-Spolaor}, S. 2013, Classical and Quantum Gravity, 30, 224013,
  \dodoi{10.1088/0264-9381/30/22/224013}

\bibitem[{{Detweiler}(1979)}]{Detweiler1979}
{Detweiler}, S. 1979, \apj, 234, 1100, \dodoi{10.1086/157593}

\bibitem[{{Drake} {et~al.}(2009){Drake}, {Djorgovski}, {Mahabal}, {Beshore},
  {Larson}, {Graham}, {Williams}, {Christensen}, {Catelan}, \&
  {Boattini}}]{2009ApJ...696..870D}
{Drake}, A.~J., {Djorgovski}, S.~G., {Mahabal}, A., {et~al.} 2009, \apj, 696,
  870, \dodoi{10.1088/0004-637X/696/1/870}

\bibitem[{{Ellis} {et~al.}(2012){Ellis}, {Siemens}, \&
  {Creighton}}]{2012ApJ...756..175E}
{Ellis}, J.~A., {Siemens}, X., \& {Creighton}, J.~D.~E. 2012, \apj, 756, 175,
  \dodoi{10.1088/0004-637X/756/2/175}

\bibitem[{{Feng} {et~al.}(2019){Feng}, {Li}, {Li}, \&
  {Wang}}]{2019arXiv190703460F}
{Feng}, Y., {Li}, D., {Li}, Y.-R., \& {Wang}, J.-M. 2019, arXiv e-prints,
  arXiv:1907.03460.
\newblock \doarXiv{1907.03460}

\bibitem[{{Foster} \& {Backer}(1990)}]{Foster_Backer90}
{Foster}, R.~S., \& {Backer}, D.~C. 1990, \apj, 361, 300,
  \dodoi{10.1086/169195}

\bibitem[{{Genel} {et~al.}(2014){Genel}, {Vogelsberger}, {Springel}, {Sijacki},
  {Nelson}, {Snyder}, {Rodriguez-Gomez}, {Torrey}, \&
  {Hernquist}}]{2014MNRAS.445..175G}
{Genel}, S., {Vogelsberger}, M., {Springel}, V., {et~al.} 2014, \mnras, 445,
  175, \dodoi{10.1093/mnras/stu1654}

\bibitem[{{Graham} {et~al.}(2015{\natexlab{a}}){Graham}, {Djorgovski}, {Stern},
  {Glikman}, {Drake}, {Mahabal}, {Donalek}, {Larson}, \&
  {Christensen}}]{2015Natur.518...74G}
{Graham}, M.~J., {Djorgovski}, S.~G., {Stern}, D., {et~al.} 2015{\natexlab{a}},
  \nat, 518, 74, \dodoi{10.1038/nature14143}

\bibitem[{{Graham} {et~al.}(2015{\natexlab{b}}){Graham}, {Djorgovski}, {Stern},
  {Drake}, {Mahabal}, {Donalek}, {Glikman}, {Larson}, \&
  {Christensen}}]{2015MNRAS.453.1562G}
---. 2015{\natexlab{b}}, \mnras, 453, 1562, \dodoi{10.1093/mnras/stv1726}

\bibitem[{{Hobbs} {et~al.}(2019){Hobbs}, {Dai}, {Manchester}, {Shannon},
  {Kerr}, {Lee}, \& {Xu}}]{2019RAA....19...20H}
{Hobbs}, G., {Dai}, S., {Manchester}, R.~N., {et~al.} 2019, Research in
  Astronomy and Astrophysics, 19, 020, \dodoi{10.1088/1674-4527/19/2/20}

\bibitem[{{Iguchi} {et~al.}(2010){Iguchi}, {Okuda}, \& {Sudou}}]{Iguchi2010}
{Iguchi}, S., {Okuda}, T., \& {Sudou}, H. 2010, \apjl, 724, L166,
  \dodoi{10.1088/2041-8205/724/2/L166}

\bibitem[{{Janssen} {et~al.}(2015){Janssen}, {Hobbs}, {McLaughlin}, {Bassa},
  {Deller}, {Kramer}, {Lee}, {Mingarelli}, {Rosado}, {Sanidas}, {Sesana},
  {Shao}, {Stairs}, {Stappers}, \& {Verbiest}}]{2015aska.confE..37J}
{Janssen}, G., {Hobbs}, G., {McLaughlin}, M., {et~al.} 2015, in Advancing
  Astrophysics with the Square Kilometre Array (AASKA14), 37.
\newblock \doarXiv{1501.00127}

\bibitem[{{Jaranowski} {et~al.}(1998){Jaranowski}, {Kr{\'o}lak}, \&
  {Schutz}}]{1998PhRvD..58f3001J}
{Jaranowski}, P., {Kr{\'o}lak}, A., \& {Schutz}, B.~F. 1998, \prd, 58, 063001,
  \dodoi{10.1103/PhysRevD.58.063001}

\bibitem[{{Jarrett} {et~al.}(2000){Jarrett}, {Chester}, {Cutri}, {Schneider},
  {Skrutskie}, \& {Huchra}}]{2000AJ....119.2498J}
{Jarrett}, T.~H., {Chester}, T., {Cutri}, R., {et~al.} 2000, \aj, 119, 2498,
  \dodoi{10.1086/301330}

\bibitem[{{Kelley} {et~al.}(2019){Kelley}, {Charisi}, {Burke-Spolaor}, {Simon},
  {Blecha}, {Bogdanovic}, {Colpi}, {Comerford}, {D'Orazio}, \&
  {Dotti}}]{2019BAAS...51c.490K}
{Kelley}, L., {Charisi}, M., {Burke-Spolaor}, S., {et~al.} 2019, in \baas,
  Vol.~51, 490.
\newblock \doarXiv{1903.07644}

\bibitem[{{Li} {et~al.}(2016){Li}, {Wang}, {Ho}, {Lu}, {Qiu}, {Du}, {Hu},
  {Huang}, {Zhang}, {Wang}, \& {Bai}}]{2016ApJ...822....4L}
{Li}, Y.-R., {Wang}, J.-M., {Ho}, L.~C., {et~al.} 2016, \apj, 822, 4,
  \dodoi{10.3847/0004-637X/822/1/4}

\bibitem[{{Li} {et~al.}(2019){Li}, {Wang}, {Zhang}, {Wang}, {Huang}, {Lu},
  {Hu}, {Du}, {Bon}, {Ho}, {Bai}, {Bian}, {Yuan}, {Winkler}, {Denissyuk},
  {Valiullin}, {Bon}, \& {Popovi{\'c}}}]{2019ApJS..241...33L}
{Li}, Y.-R., {Wang}, J.-M., {Zhang}, Z.-X., {et~al.} 2019, \apjs, 241, 33,
  \dodoi{10.3847/1538-4365/ab0ec5}

\bibitem[{{LIGO Scientific Collaboration} {et~al.}(2015){LIGO Scientific
  Collaboration}, {Aasi}, {Abbott}, {Abbott}, {Abbott}, {Abernathy}, {Ackley},
  {Adams}, {Adams}, \& {Addesso}}]{2015CQGra..32g4001L}
{LIGO Scientific Collaboration}, {Aasi}, J., {Abbott}, B.~P., {et~al.} 2015,
  Classical and Quantum Gravity, 32, 074001,
  \dodoi{10.1088/0264-9381/32/7/074001}

\bibitem[{{Manchester} \& {IPTA}(2013)}]{2013CQGra..30v4010M}
{Manchester}, R.~N., \& {IPTA}. 2013, Classical and Quantum Gravity, 30,
  224010, \dodoi{10.1088/0264-9381/30/22/224010}

\bibitem[{{Manchester} {et~al.}(2013){Manchester}, {Hobbs}, {Bailes}, {Coles},
  {van Straten}, {Keith}, {Shannon}, {Bhat}, {Brown}, {Burke-Spolaor},
  {Champion}, {Chaudhary}, {Edwards}, {Hampson}, {Hotan}, {Jameson}, {Jenet},
  {Kesteven}, {Khoo}, {Kocz}, {Maciesiak}, {Oslowski}, {Ravi}, {Reynolds},
  {Sarkissian}, {Verbiest}, {Wen}, {Wilson}, {Yardley}, {Yan}, \&
  {You}}]{2013PASA...30...17M}
{Manchester}, R.~N., {Hobbs}, G., {Bailes}, M., {et~al.} 2013, \apj, 30, e017,
  \dodoi{10.1017/pasa.2012.017}

\bibitem[{{McConnell} \& {Ma}(2013)}]{2013ApJ...764..184M}
{McConnell}, N.~J., \& {Ma}, C.-P. 2013, \apj, 764, 184,
  \dodoi{10.1088/0004-637X/764/2/184}

\bibitem[{Mingarelli(2017)}]{chiara_mingarelli_2017_838712}
Mingarelli, C. 2017, ChiaraMingarelli/nanohertz\_GWs: First release!,
  \dodoi{10.5281/zenodo.838712}

\bibitem[{{Mingarelli} {et~al.}(2012){Mingarelli}, {Grover}, {Sidery}, {Smith},
  \& {Vecchio}}]{2012PhRvL.109h1104M}
{Mingarelli}, C.~M.~F., {Grover}, K., {Sidery}, T., {Smith}, R.~J.~E., \&
  {Vecchio}, A. 2012, Physical Review Letters, 109, 081104,
  \dodoi{10.1103/PhysRevLett.109.081104}

\bibitem[{{Mingarelli} {et~al.}(2017){Mingarelli}, {Lazio}, {Sesana}, {Greene},
  {Ellis}, {Ma}, {Croft}, {Burke-Spolaor}, \& {Taylor}}]{2017NatAs...1..886M}
{Mingarelli}, C.~M.~F., {Lazio}, T.~J.~W., {Sesana}, A., {et~al.} 2017, Nature
  Astronomy, 1, 886, \dodoi{10.1038/s41550-017-0299-6}

\bibitem[{{Muzzin} {et~al.}(2013){Muzzin}, {Marchesini}, {Stefanon}, {Franx},
  {McCracken}, {Milvang-Jensen}, {Dunlop}, {Fynbo}, {Brammer}, {Labb{\'e}}, \&
  {van Dokkum}}]{2013ApJ...777...18M}
{Muzzin}, A., {Marchesini}, D., {Stefanon}, M., {et~al.} 2013, \apj, 777, 18,
  \dodoi{10.1088/0004-637X/777/1/18}

\bibitem[{{Nan} {et~al.}(2011){Nan}, {Li}, {Jin}, {Wang}, {Zhu}, {Zhu},
  {Zhang}, {Yue}, \& {Qian}}]{2011IJMPD..20..989N}
{Nan}, R., {Li}, D., {Jin}, C., {et~al.} 2011, International Journal of Modern
  Physics D, 20, 989, \dodoi{10.1142/S0218271811019335}

\bibitem[{{Peters}(1964)}]{1964PhRv..136.1224P}
{Peters}, P.~C. 1964, Physical Review, 136, 1224,
  \dodoi{10.1103/PhysRev.136.B1224}

\bibitem[{{Planck Collaboration} {et~al.}(2016){Planck Collaboration}, {Ade},
  {Aghanim}, {Arnaud}, {Ashdown}, {Aumont}, {Baccigalupi}, {Banday},
  {Barreiro}, \& {Bartlett}}]{2016A&A...594A..13P}
{Planck Collaboration}, {Ade}, P.~A.~R., {Aghanim}, N., {et~al.} 2016, \aap,
  594, A13, \dodoi{10.1051/0004-6361/201525830}

\bibitem[{{Porayko} {et~al.}(2018){Porayko}, {Zhu}, {Levin}, {Hui}, {Hobbs},
  {Grudskaya}, {Postnov}, {Bailes}, {Bhat}, \& {Coles}}]{2018PhRvD..98j2002P}
{Porayko}, N.~K., {Zhu}, X., {Levin}, Y., {et~al.} 2018, \prd, 98, 102002,
  \dodoi{10.1103/PhysRevD.98.102002}

\bibitem[{{Ravi} {et~al.}(2015){Ravi}, {Wyithe}, {Shannon}, \&
  {Hobbs}}]{2015MNRAS.447.2772R}
{Ravi}, V., {Wyithe}, J.~S.~B., {Shannon}, R.~M., \& {Hobbs}, G. 2015, \mnras,
  447, 2772, \dodoi{10.1093/mnras/stu2659}

\bibitem[{{Rodriguez-Gomez} {et~al.}(2015){Rodriguez-Gomez}, {Genel},
  {Vogelsberger}, {Sijacki}, {Pillepich}, {Sales}, {Torrey}, {Snyder},
  {Nelson}, \& {Springel}}]{2015MNRAS.449...49R}
{Rodriguez-Gomez}, V., {Genel}, S., {Vogelsberger}, M., {et~al.} 2015, \mnras,
  449, 49, \dodoi{10.1093/mnras/stv264}

\bibitem[{{Sesana}(2013)}]{2013MNRAS.433L...1S}
{Sesana}, A. 2013, \mnras, 433, L1, \dodoi{10.1093/mnrasl/slt034}

\bibitem[{{Sesana} {et~al.}(2018){Sesana}, {Haiman}, {Kocsis}, \&
  {Kelley}}]{2018ApJ...856...42S}
{Sesana}, A., {Haiman}, Z., {Kocsis}, B., \& {Kelley}, L.~Z. 2018, \apj, 856,
  42, \dodoi{10.3847/1538-4357/aaad0f}

\bibitem[{{Sesana} \& {Khan}(2015)}]{2015MNRAS.454L..66S}
{Sesana}, A., \& {Khan}, F.~M. 2015, \mnras, 454, L66,
  \dodoi{10.1093/mnrasl/slv131}

\bibitem[{{Skrutskie} {et~al.}(2006){Skrutskie}, {Cutri}, {Stiening},
  {Weinberg}, {Schneider}, {Carpenter}, {Beichman}, {Capps}, {Chester}, \&
  {Elias}}]{2006AJ....131.1163S}
{Skrutskie}, M.~F., {Cutri}, R.~M., {Stiening}, R., {et~al.} 2006, \aj, 131,
  1163, \dodoi{10.1086/498708}

\bibitem[{{Smits} {et~al.}(2009){Smits}, {Kramer}, {Stappers}, {Lorimer},
  {Cordes}, \& {Faulkner}}]{2009A&A...493.1161S}
{Smits}, R., {Kramer}, M., {Stappers}, B., {et~al.} 2009, \aap, 493, 1161,
  \dodoi{10.1051/0004-6361:200810383}

\bibitem[{{Valtonen} {et~al.}(2016){Valtonen}, {Zola}, {Ciprini}, {Gopakumar},
  {Matsumoto}, {Sadakane}, {Kidger}, {Gazeas}, {Nilsson}, {Berdyugin},
  {Piirola}, {Jermak}, {Baliyan}, {Alicavus}, {Boyd}, {Campas Torrent},
  {Campos}, {Carrillo G{\'o}mez}, {Caton}, {Chavushyan}, {Dalessio}, {Debski},
  {Dimitrov}, {Drozdz}, {Er}, {Erdem}, {Escartin P{\'e}rez}, {Fallah Ramazani},
  {Filippenko}, {Ganesh}, {Garcia}, {G{\'o}mez Pinilla}, {Gopinathan},
  {Haislip}, {Hudec}, {Hurst}, {Ivarsen}, {Jelinek}, {Joshi}, {Kagitani},
  {Kaur}, {Keel}, {LaCluyze}, {Lee}, {Lindfors}, {Lozano de Haro}, {Moore},
  {Mugrauer}, {Naves Nogues}, {Neely}, {Nelson}, {Ogloza}, {Okano}, {Pandey},
  {Perri}, {Pihajoki}, {Poyner}, {Provencal}, {Pursimo}, {Raj}, {Reichart},
  {Reinthal}, {Sadegi}, {Sakanoi}, {Salto Gonz{\'a}lez}, {Sameer}, {Schweyer},
  {Siwak}, {Sold{\'a}n Alfaro}, {Sonbas}, {Steele}, {Stocke}, {Strobl},
  {Takalo}, {Tomov}, {Tremosa Espasa}, {Valdes}, {Valero P{\'e}rez},
  {Verrecchia}, {Webb}, {Yoneda}, {Zejmo}, {Zheng}, {Telting}, {Saario},
  {Reynolds}, {Kvammen}, {Gafton}, {Karjalainen}, {Harmanen}, \&
  {Blay}}]{Valtonen2016}
{Valtonen}, M.~J., {Zola}, S., {Ciprini}, S., {et~al.} 2016, \apjl, 819, L37,
  \dodoi{10.3847/2041-8205/819/2/L37}

\bibitem[{{Wang} \& {Mohanty}(2017)}]{2017PhRvL.118o1104W}
{Wang}, Y., \& {Mohanty}, S.~D. 2017, Physical Review Letters, 118, 151104,
  \dodoi{10.1103/PhysRevLett.118.151104}

\bibitem[{{Yan} {et~al.}(2015){Yan}, {Lu}, {Dai}, \&
  {Yu}}]{2015ApJ...809..117Y}
{Yan}, C.-S., {Lu}, Y., {Dai}, X., \& {Yu}, Q. 2015, \apj, 809, 117,
  \dodoi{10.1088/0004-637X/809/2/117}

\end{thebibliography}

\end{document}